\documentclass[sigconf]{acmart}

\AtBeginDocument{%
  }

\setcopyright{acmlicensed}
\copyrightyear{2024}
\acmYear{2024}
\acmDOI{10.1145/3665318.3677149}

\acmConference[WEB3D '24]{The 29th International ACM Conference on 3D Web Technology}{September 25--27, 2024}{Guimarães, Portugal}
\acmBooktitle{The 29th International ACM Conference on 3D Web Technology (WEB3D '24), September 25--27, 2024, Guimarães, Portugal}
\acmISBN{979-8-4007-0689-9/24/09}

\usepackage{algorithm}% http://ctan.org/pkg/algorithm
\usepackage{algpseudocode}% http://ctan.org/pkg/algorithmicx
\usepackage{subfigure}

\begin{document}

\title{Holonomy: A Virtual Environment based on Hyperbolic Space}


\author{Martin Skrodzki}
\affiliation{%
  \institution{Comp.~Graphics and Vis., TU Delft}
  \city{Delft}
  \country{Netherlands}}
\email{0000-0002-8126-0511}
\authornote{Corresponding author: \href{mailto:mail@ms-math-computer.science}{mail@ms-math-computer.science}}

\author{Scott Jochems}
\affiliation{%
  \institution{TU Delft}
  \city{Delft}
  \country{Netherlands}}
\email{0009-0000-2716-2112}

\author{Joris Rijsdijk}
\affiliation{%
  \institution{TU Delft}
  \city{Delft}
  \country{Netherlands}}
\email{0009-0006-3767-0202}

\author{Ravi Snellenberg}
\affiliation{%
  \institution{TU Delft}
  \city{Delft}
  \country{Netherlands}}
\email{0009-0005-7571-1813}

\author{Rafael Bidarra}
\affiliation{%
  \institution{Comp.~Graphics and Vis., TU Delft}
  \city{Delft}
  \country{Netherlands}}
\email{0000-0003-4281-6019}

\renewcommand{\shortauthors}{Skrodzki, Martin, et al.}

\begin{abstract}
\emph{Holonomy} is a virtual environment based on the mathematical concept of hyperbolic geometry. 
Unlike other environments, \emph{Holonomy} allows users to seamlessly explore an infinite hyperbolic space by physically walking.
They use their body as the controller, eliminating the need for teleportation or other artificial VR locomotion methods. 
This paper discusses the development of \emph{Holonomy}, highlighting the technical challenges faced and overcome during its creation, including rendering complex hyperbolic environments, populating the space with objects, and implementing algorithms for finding shortest paths in the underlying non-Euclidean geometry.
Furthermore, we present a proof-of-concept implementation in the form of a VR navigation game and some preliminary learning outcomes from this implementation.
\end{abstract}



\begin{teaserfigure}
    \includegraphics[width=\textwidth]{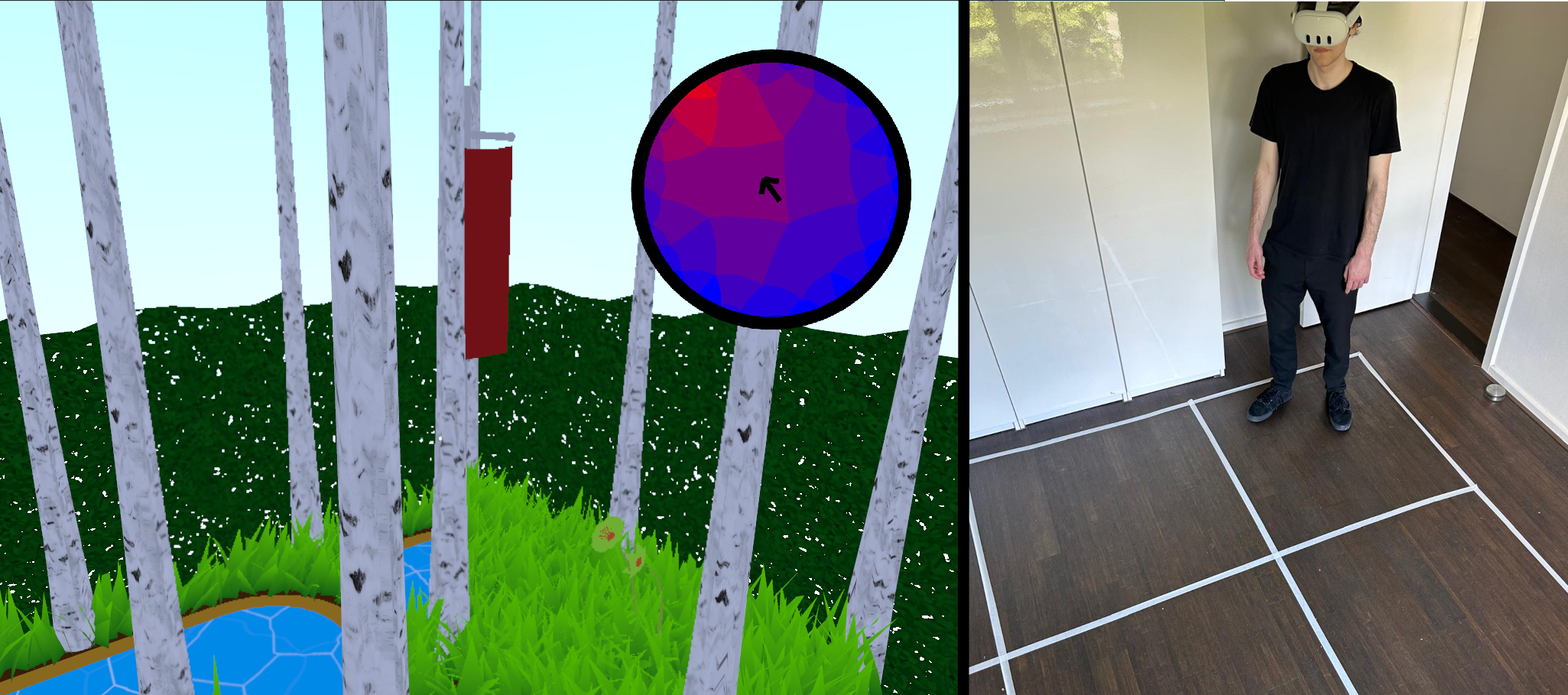}
    \caption{
        Left: Rendering of the hyperbolic virtual environment \emph{Holonomy}, including a mini-map in the HUD.
        Hyperbolic effects are visible in the flag and the creeks partially vanishing behind the trees.
        Right: A player in the physical move area, with the corresponding square tiles marked on the floor.
    }
    \Description{
        Two images side by side, on the left the virtual environment's depiction, where a flag can be barely seen together with a mini-map. The right image depicts a player with a Virtual Reality Headset in a square grid. 
    }
  \label{fig:teaser}
\end{teaserfigure}


\maketitle

\section{Introduction}
\label{sec:Introduction}

In an increasingly interconnected and digitized world, people communicate and interact in various ways and via multiple different platforms.
In this context, virtual environments rapidly gain importance for both professional and leisurely exchanges.
Recent hardware developments like Apple's \emph{Vision Pro} seek to integrate virtual content into the user's physical surroundings, merging the two in the amalgam of Spatial Computing. 
Similarly, the Metaverse movement aims to integrate everyday processes within a virtual environment.
Using virtual, artificial, and mixed reality blurs the lines between physical and virtual worlds. 
In the context of these developments, we present a virtual environment based on the mathematical concept of hyperbolic space.

Hyperbolic geometry, a non-Euclidean geometry pioneered by mathematicians such as Nikolai Lobachevsky and János Bolyai in the 19th century, offers a fascinating alternative to the familiar principles of Euclidean space. 
In hyperbolic geometry, the parallel postulate is replaced by the concept that through a given point, an infinite number of lines can be drawn parallel to a given line, leading to properties such as space being negatively curved and exhibiting unique geometrical constructions. 
Understanding and visualizing hyperbolic geometry can be challenging due to its abstract nature and departure from our intuitive understanding of space.
However, the unique geometric properties of hyperbolic spaces provide creative opportunities for creating new virtual environments.

Hyperbolic spaces have been used very successfully, specifically regarding data representation.
For example, it is possible to embed hierarchical structures like trees into two-dimensional hyperbolic space with arbitrarily low distortion~\cite{sarkar2011low}.
By this property, previous works have successfully embedded social networks~\cite{verbeek2014metric} or the Internet~\cite{boguna2010sustaining} into hyperbolic space.
This usage as an exploration space for virtual data further motivates our creation of a new virtual environment based on hyperbolic spaces.

The advent of virtual reality (VR) technology has opened up new avenues for exploring and experiencing hyperbolic geometry in immersive and interactive ways. 
VR provides a platform where users can not only passively observe hyperbolic spaces but actively engage with and navigate through them. 
We apply specific terminology in the following discussion on navigating VR and throughout the paper.
The \emph{physical space} is the real-world space the user occupies and where they perform their movements, see Figure~\ref{fig:teaser}, right.
In contrast, the \emph{virtual space} refers to the virtual environment in which the user is immersed, the virtual world rendered via the VR headset, see Figure~\ref{fig:teaser}, left.
Within the physical space, the user has a dedicated \emph{move area}: ideally, a square shape, physically unobstructed free space, indicated on the floor in Figure~\ref{fig:teaser}, right.
As the user moves through this area physically, we refer to it as \emph{walking}.
Every walking motion will be translated into movement in the virtual environment.
Those parts of the virtual environment shown to the user will be called \emph{accessible}, while hidden areas are \emph{inaccessible}.

There are several ways of allowing users to navigate through the virtual environment. 
One option is to allow them to walk within the move area. 
The benefit of this method is the intuitive understanding and freedom of movement. 
This method is not without flaws, as it either requires expensive equipment---such as specialized treadmills~\cite{iwata1999walking} or a CAVE\textregistered~\cite{razzaque2002redirected}---or the use of redirected walking techniques in a large move area~\cite{fan2022redirected} to give users the required freedom of movement.
Aside from the user's walking, the most common way is some form of controller input to glide or teleport the user, which in turn can create VR sickness~\cite{monteiro2021evaluating}.
The benefit of this method is that it requires only a minuscule move area to work. 
However, it reduces the feeling of presence and thus immersion within a virtual environment~\cite{clifton2020effects}.
In contrast, the virtual environment presented in this paper provides an infinitely explorable space that can be traversed solely by walking in a relatively small move area without additional hardware requirements.

Creating a virtual environment for VR comes with its own set of crucial technical challenges compared to a 3D environment explored via a display.
The virtual environment must perform at a stable high frame rate, high resolution, and very low input latency~\cite{wang2023effect}. 
To achieve this on widely available consumer-level hardware implies a sacrifice in visual quality, paired with a lower capacity for overall scene complexity. 
This is an additional challenge in a virtual space based on hyperbolic geometry, as classical rendering techniques, pipelines, and hardware are optimized for processing Euclidean scenes.
The same holds for representations of the environment and their object population, as well as navigational cues for the user, for instance via the display of the shortest path to a point of interest. 
In this paper, we tackle these challenges and offer some solutions.
Note that while we developed VR first, our environment does not exclude other modes of exploration.
Users can also explore the environment with a traditional screen and controller setup.

In summary, this paper introduces \emph{Holonomy}, a new virtual environment based on principles of hyperbolic geometry.
We discuss its development choices, technical challenges, and potential applications. 
Through the fusion of hyperbolic geometry and VR technology, Holonomy offers a novel approach to virtual exploration in non-Euclidean spaces.
The specific contributions of this work are:
\begin{itemize}
    \item A virtual environment, based on the mathematical concept of hyperbolic spaces, Section~\ref{sec:HolonomyTheVirtualEnvironment}.
    \item Potentially infinite exploring of a virtual world solely by physically walking in a move area, without the need for teleportation or other artificial locomotion, Section~\ref{sec:HolonomyTheVirtualEnvironment}.
    \item Efficient rendering techniques for the virtual environment to maintain high frame rates suitable for VR, Section~\ref{sec:Rendering}.
    \item Adaptation of a wave function collapse algorithm to populate the virtual environment, Section~\ref{sec:PopulatingTheVirtualSpaceWithObjects}.
    \item Adjustment of shortest path algorithms for user guidance in navigating the environment, Section~\ref{sec:Navigation}.
    \item A proof-of-concept implementation of our environment as a VR navigation game, Section~\ref{sec:ProofOfConcept}.
\end{itemize}
We see two immediate applications of our work.
First, the interactive teaching of aspects of hyperbolic geometry.
Second, usage of the system for the presentation of unique navigational challenges, for instance, within cognition studies. 
We will elaborate on these applications in Section~\ref{sec:ConclusionsImpactAndOutlook}.

\section{Background}

\begin{figure}
    \centering
    \includegraphics[width=1.\linewidth]{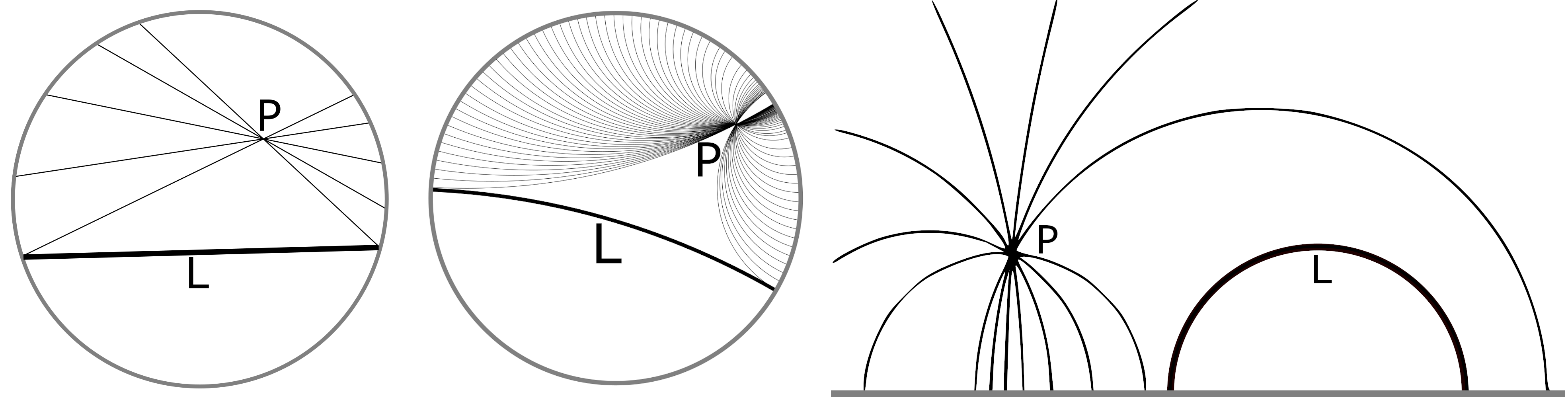}
    \caption{
        A line $L$ and several parallels to $L$ through a point $P$ in different models
of the hyperbolic plane. 
        Left:~Beltrami-Klein model in a disk, straight hyperbolic lines remain straight; 
        Center:~Poincaré disk model, straight hyperbolic lines become curved; 
        Right:~Poincaré half-plane model, straight hyperbolic lines are either vertical or half circles, but angles are kept intact.
        Reproduced with permission~\cite{skrodzki2021illustrations}.
    }
    \Description{
        On the left a circle is shown, slightly below the center of this circle a straight horizontal line labeled as L is drawn. Slightly above and to the right of the center a point labeled as P is drawn. 4 straight lines drawn at different angles go through this point P. In the middle, a circle is drawn with a curved line labeled L slightly below the center. A point P is drawn slightly above and to the right of the center. An indicated "infinite" amount of curved lines go through this point P. On the right, a horizontal line is drawn. On the right side of this line, a half circle is drawn labeled L. To the top left of L a point P is drawn. Through this point, P 7 curved lines are drawn. 2 of which complete a half circle on the horizontal line. One of these half circles is smaller than L and completes to the left of L. The other is bigger and arches over L.
    }
    \label{fig:2DHyperbolicSpace}
\end{figure}

Before discussing our virtual environment, we set the stage by introducing some necessary background.
First, we cover basic notions of hyperbolic geometry, highlighting the notable difference to Euclidean geometry that through a given point~$P$ off a given line~$L$, there are infinitely many lines that do not intersect~$L$, see Figure~\ref{fig:2DHyperbolicSpace}.
Second, we introduce the notion of a \emph{hyperbolic tiling}, which will provide the main mode of connection between the user's move area and the virtual environment.

\subsection{Visualizing Hyperbolic geometry}

Various visualization models have been developed to help conceptualize hyperbolic space. 
The Beltrami-Klein model, for instance, renders the hyperbolic plane into a disk in Euclidean space, see Figure~\ref{fig:2DHyperbolicSpace}, left.
It maps straight hyperbolic lines to straight lines in the model but distorts their angles.
The namesakes are Eugenio Beltrami and Felix Klein.
Another common approach is the Poincaré disk model, named after the French mathematician Henri Poincaré.
In this model, hyperbolic space is again represented within a Euclidean disk but straight hyperbolic lines are rendered as circular arcs orthogonal to the disk's boundary, see Figure~\ref{fig:2DHyperbolicSpace}, center. 
Another visualization model is the Poincaré half-plane model, which utilizes the entire upper half of the Euclidean plane to map the hyperbolic plane. 
It renders straight hyperbolic lines as vertical lines or half circles while preserving angles, see Figure~\ref{fig:2DHyperbolicSpace}, right.
All three models offer distinct perspectives on hyperbolic geometry, aiding comprehension and study of its unique properties and structures~\cite[Chapter~3.6]{kinsey2011geometry}.

Maurits Cornelis Escher, the renowned Dutch artist, played a pivotal role in popularizing the Poincaré disk model through his iconic \emph{Circle Limit} series. 
Inspired by a diagram, which the British mathematician Donald Coxeter used in a 1957 mathematical paper on hyperbolic geometry~\cite{coxeter1957crystal}, Escher created intricate woodcut prints depicting tessellations of hyperbolic space, where angels, demons, and other figures form repeating patterns that fill the entire plane. 
Tesselations, also called tilings, of the hyperbolic plane are the foundation for the following discussion.
For a more general introduction to hyperbolic tilings, see~\cite{Conway-symmetries}.

Note that a hyperbolic space is a homogeneous space with constant negative curvature.
This property is exhibited for instance in the fact that the sum of angles of a triangle is less than~$180$ degrees. 
Another way to express this fact via tilings is to count the number of tiles that meet at a vertex.
Consider an ordinary bathroom tiling of square tiles, see Figure~\ref{fig:HyperbolicViewDifference}, left.
Here, four squares meet at a vertex, giving an angle sum of~$360$ degrees.
In a corresponding hyperbolic square tiling, more than four squares can meet, see Figure~\ref{fig:HyperbolicViewDifference}, right, for an example of five squares meeting.

\begin{figure*}
    \centering
    \includegraphics[width=1.\textwidth]{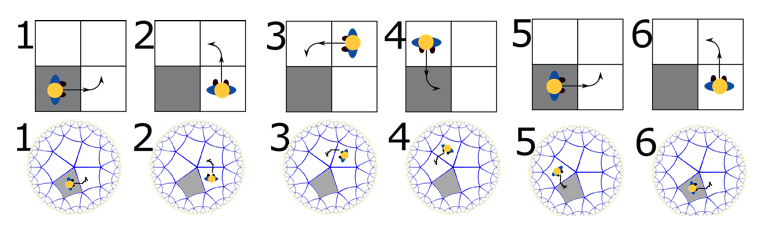}
    \caption{
        Top row: A user moving their body through a physical~$2\times2$ Euclidean square grid. 
        Bottom row: The movements are mapped to a square tiling in the hyperbolic plane, with five squares meeting at a vertex.
        Reproduced with permission~\cite{jochems2023mini-map}.
    }
    \Description{
        This image consists of 2 rows. The top row shows 6 $2\times2$ squares each labeled 1 to 6. The bottom left square in each is dark grey while the others are all white. A character can be seen walking through these squares with an arrow in front indicating the next step. The character starts in the bottom left and then goes counterclockwise around the center with each consecutive grid, finally in grid number 6 ending up in the bottom right. The bottom row shows 6 hyperbolic disks where 5 areas meet. The bottom area is colored dark grey. each disk is labeled 1 to 6. A character can be seen walking through these areas with an arrow in front indicating the next step. The character starts in the dark grey area and moves counterclockwise around the center with each consecutive disk, finally in disk number 6 ending in the dark grey area.
    }
    \label{fig:HyperbolicWalker}
\end{figure*}

Having more than four squares meet at a point enables the effect of \emph{Holonomy}---after which our project is named---by which a closed circular walk in the physical space causes a rotation or displacement in the hyperbolic virtual space.
This was first presented as a phenomenon in VR by Henry Segerman\footnote{See the corresponding video \url{https://www.youtube.com/watch?v=ztsi0CLxmjw}.} as follows:
A user walks in a~$2\times2$ Euclidean square grid, marked on the floor of their move area.
Here, they have to take a step and perform a 90-degree left turn exactly four times to return to the square where and orientation in which they started, see Figure~\ref{fig:HyperbolicWalker}, top row.
Simultaneously, they see themselves navigating a hyperbolic square tiling in VR with five squares meeting at a vertex.
That is, after four consecutive steps and turns, they are not yet back to the starting square, see Figure~\ref{fig:HyperbolicWalker}, bottom row.
It takes them another step and a 90-degree turn to return to their virtual starting position, causing a discrepancy between their physical and virtual presence.
This discrepancy will allow for an infinite exploration within our virtual environment solely based on the user walking within their move area.

\begin{figure}
    \centering
    \includegraphics[width=0.6\linewidth]{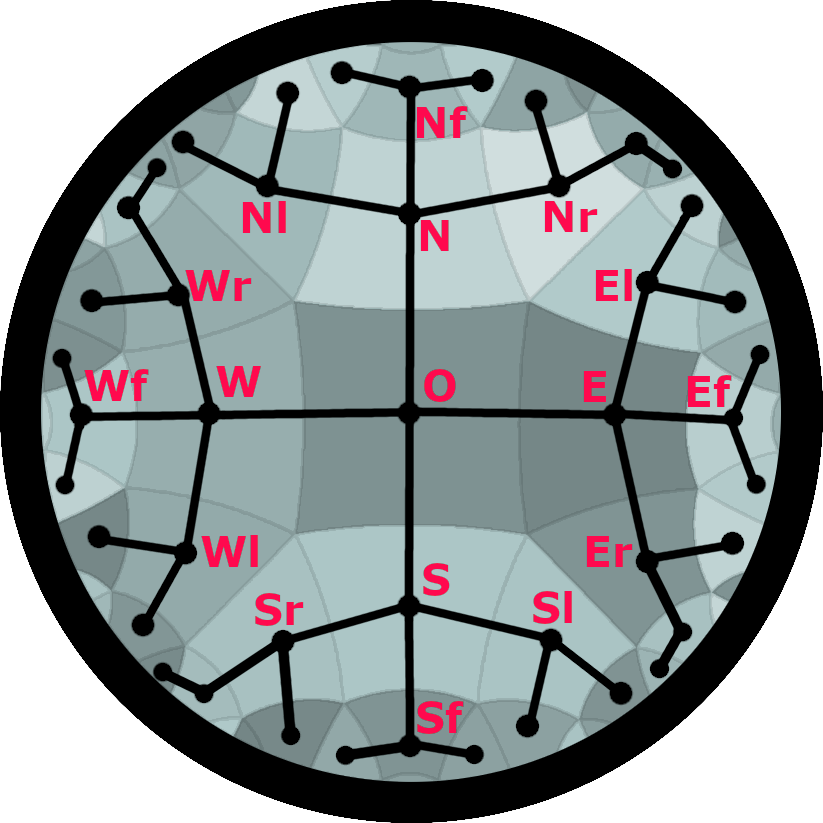}
    \caption{
        The spanning tree of the graph.
        Nodes represent the hyperbolic squares underneath.
        Labels indicate the indices of the nodes.
    }
    \Description{A circle is shown with a graph inside of it. At the center, a black dot is labeled with a red O. At each cardinal direction a black line is drawn connecting to another black dot labeled in red with its corresponding direction: N, E, S, W. From each of these black dots 3 more black lines are drawn connecting to 3 black dots each labeled in red with the corresponding cardinal direction and one of the following: l for left, f for forward and r for right depending on the direction taken from the previous black dot. Adding l, f, or r continues until the circle's border is reached. Each black dot is in the center of an area, the areas are drawn with grey lines. At each corner of an area, a total of 5 areas connect. These corners are marked with lime green dots.
    }
    \label{fig:spanning}
\end{figure}

\subsection{Representing hyperbolic tilings}
\label{sec:RepresentingHyperbolicTilings}

In our virtual environment, we will focus on the hyperbolic tiling where five squares meet at any vertex.
See Section~\ref{sec:HolonomyTheVirtualEnvironment} for reasoning regarding this design choice.
This hyperbolic tiling has to be represented by a suitable data structure.
This is a vital design choice because it immediately determines the difficulty of implementing other features such as a map of the environment. 
The representation used throughout this paper has been presented in previous work~\cite{kopczynska2021generating} and we give a brief overview here.

First, note that the tiling can be reduced to a graph representation with each square represented by a node, and two nodes connected if their squares are adjacent.
A spanning tree can be built on this graph, see Figure~\ref{fig:spanning}.
As the tree structure guarantees a unique shortest path to any node, this allows an indexing of the nodes and thus easy access to a square via the index of its node.
A discrete representation like this is preferred, because numeric coordinate representation in one of the two-dimensional hyperbolic models of Figure~\ref{fig:2DHyperbolicSpace} introduces numeric precision issues.

Following~\cite{kopczynska2021generating}, the representation of a step sequence is as follows: squares are first indexed by a choice of branch from the starting point: North, West, South, or East.
This is followed by a series of directions: Forward, Left, or Right.
Here, moving backward corresponds to deleting the last step, see Figure~\ref{fig:spanning}. 
As no shortest path includes a backward step and the squares are indexed by their respective shortest path, we exclude a specific backward step.
However, this means, that the first step needs a different representation since there are four valid directions of movement from the origin rather than three for all subsequent steps.

In building the data structure, two algorithms are used to ensure that the sequence of steps remains unique.
For instance, taking a right step from the node~$Nr$ in Figure~\ref{fig:spanning} will switch to the~$E$ branch and collapse the invalid index~$Nrr$ to the valid index~$El$, indicated in the figure by a small red arrow.

The first algorithm determines whether a step is valid. 
That is, it decides whether a step stays on the current branch of the spanning tree rather than moving to a different branch.
This decision works as follows.
A forward step is always valid. 
A right step is invalid if and only if the previous step was also a right step. 
A left step is invalid if and only if there is no right step between it and the last left step.

The second algorithm normalizes an invalid sequence of steps into a valid one, by switching to a different branch of the tree if necessary.
The normalization algorithm applies the following transformations: 
\begin{align*}
    \{x, r, r\} & \rightarrow \{r(x), l\},\\
    \{x, l, l\} & \rightarrow \{l(x), r\},\\
    \{x, \{r, f\}_n, r, r\} & \rightarrow \{r(x), l, \{f\}_n\},\\
    \{x, l, \{f\}_n, l\} & \rightarrow \{l(x), \{r, f\}_n, r\},
\end{align*}
where~$r(x)$ rotates the previous step, either branch or direction, to the right, and~$l(x)$ is a rotation to the left. 
This normalization ensures that the sequence of steps corresponds to a connected and unique series of edges of the spanning tree, starting at the point of origin. 
It allows for instance the population of the virtual environment, see Section~\ref{sec:ChallengesAndSolutions}.
\section{Related work}

\begin{figure*}
    \centering
    \subfigure[Hyperrogue]{
        \includegraphics[width=0.15\textwidth]{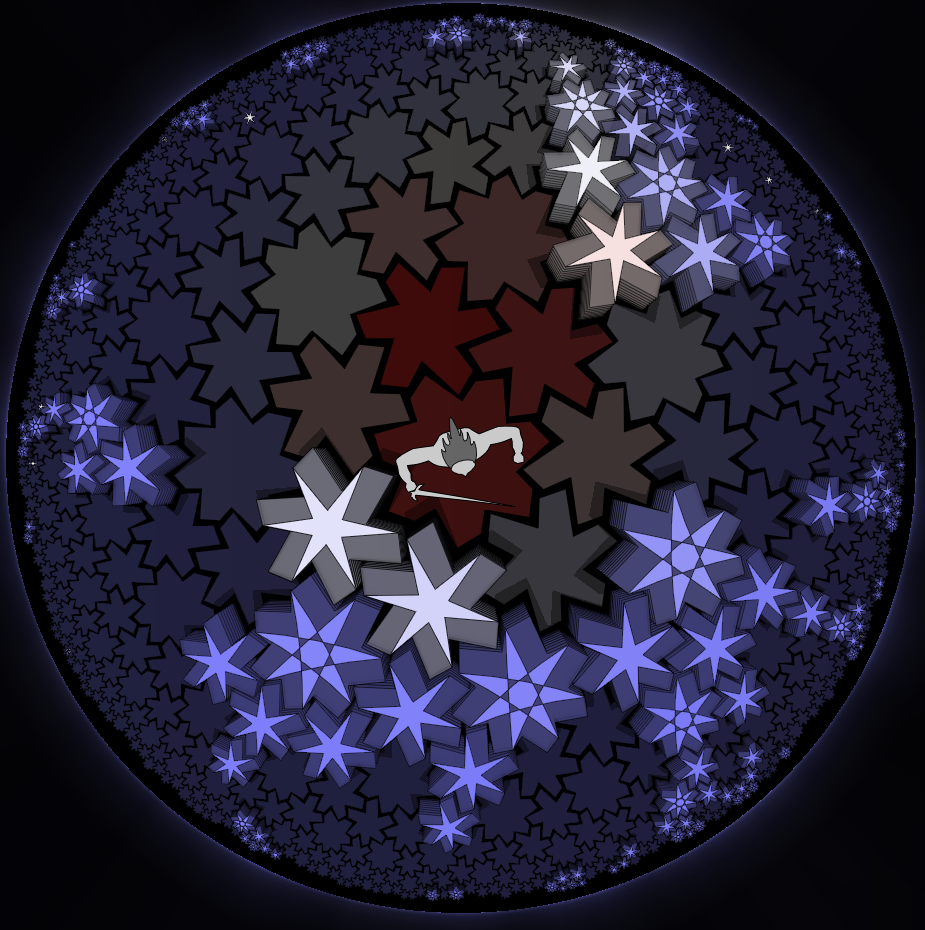}
        \label{fig:Hyperrogue}
    }
    \hfill
    \subfigure[Hyperbolica]{
        \includegraphics[width=0.15\textwidth]{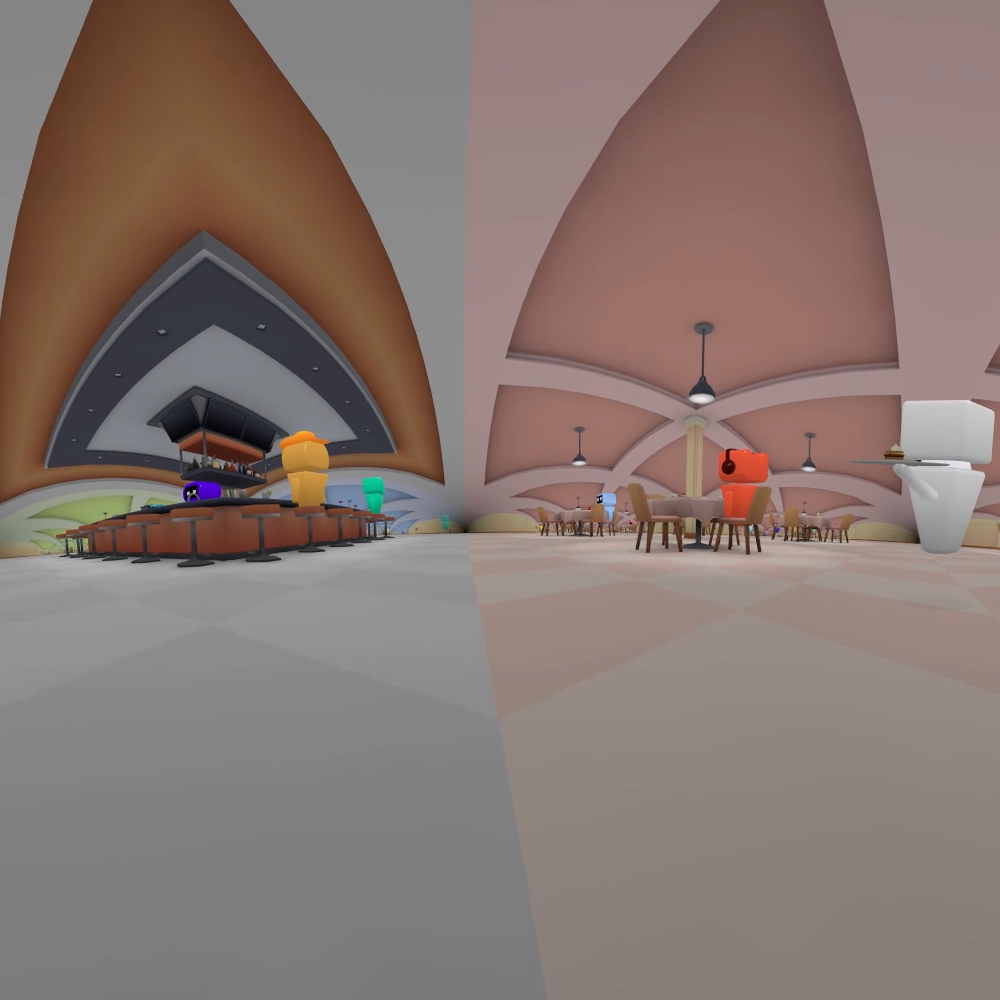}
        \label{fig:Hyperbolica}
    }
    \hfill
    \subfigure[HyperNom]{
        \includegraphics[width=0.15\textwidth]{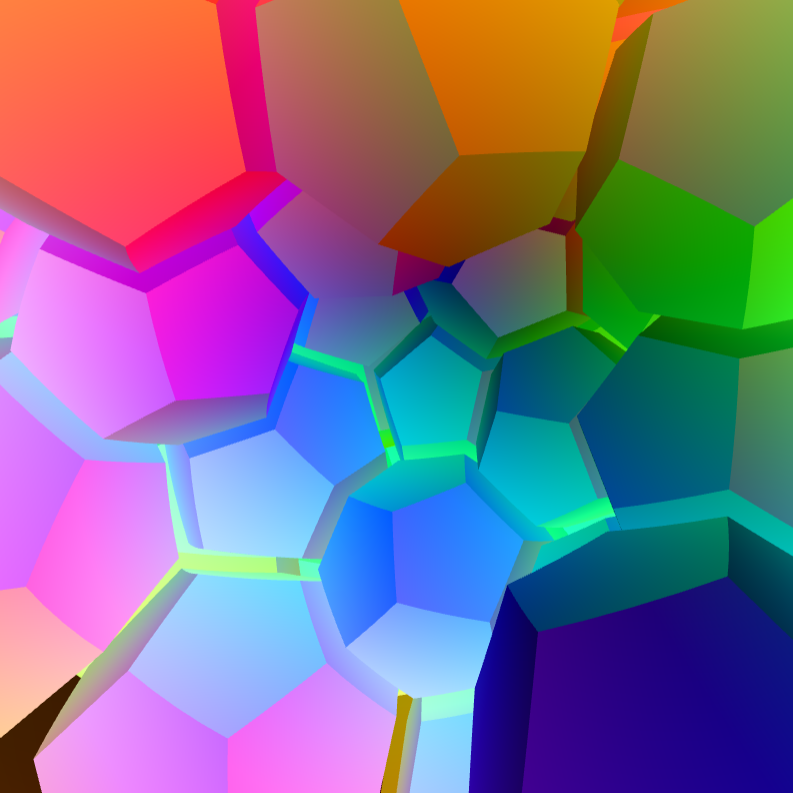}
        \label{fig:HyperNom}
    }
    \hfill
    \subfigure[HypVR]{
        \includegraphics[width=0.15\textwidth]{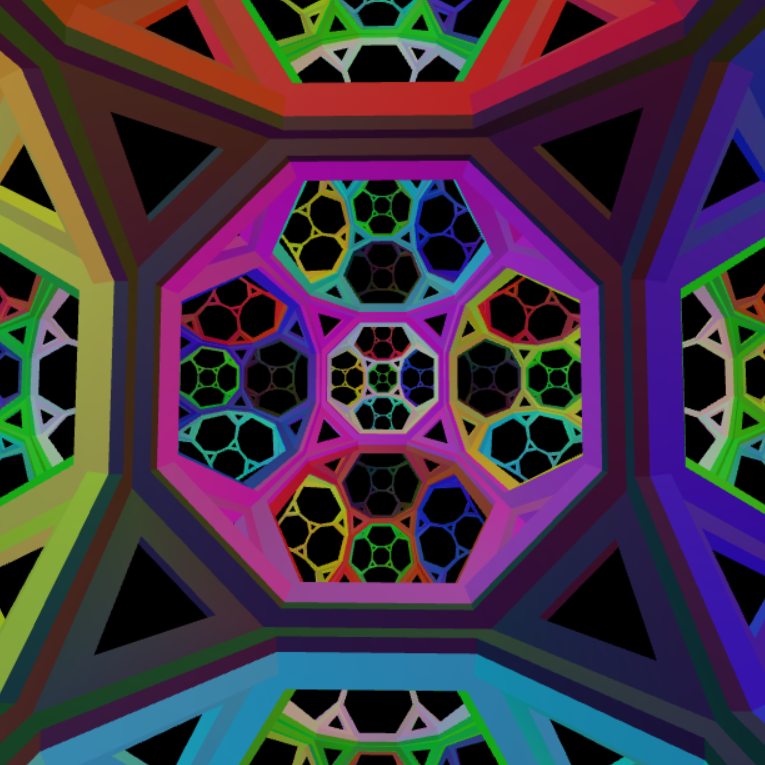}
        \label{fig:HypVR}
    }
    \hfill
    \subfigure[HypVR (RayMar.)]{
        \includegraphics[width=0.15\textwidth]{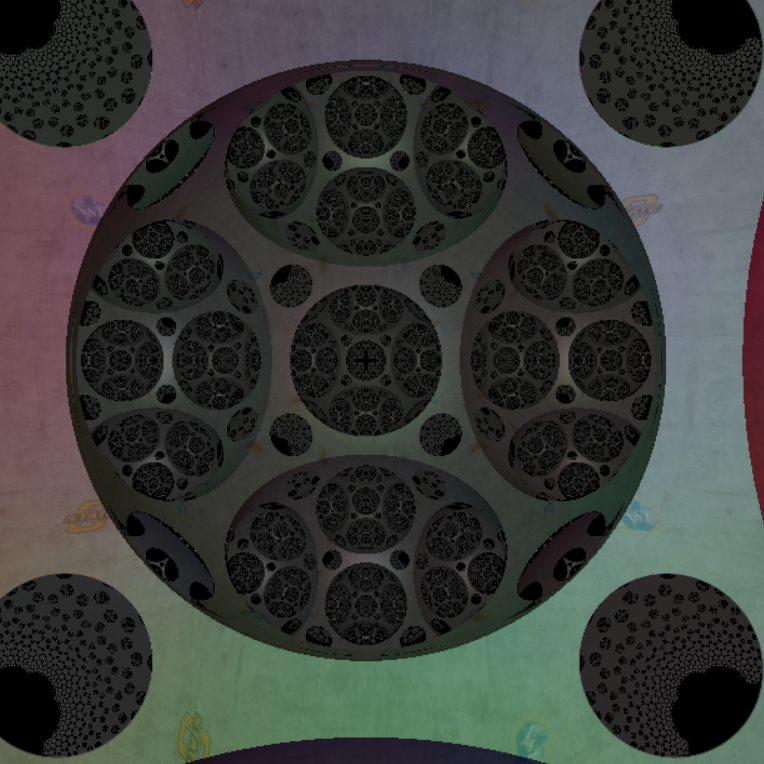}
        \label{fig:HypVRRay}
    }
    \hfill
    \subfigure[HypVR2]{
        \includegraphics[width=0.15\textwidth]{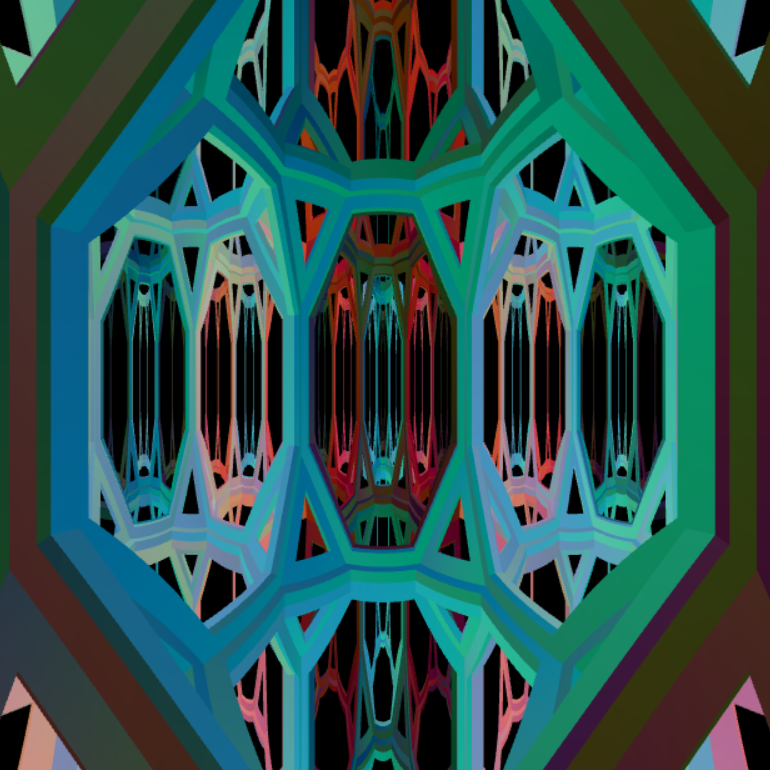}
        \label{fig:HypVR2}
    }
    \caption{
        Screenshots of VR environments, experiences, and games using hyperbolic geometry.
    }
    \Description{
        6 images are displayed in a row. The first image is a top-down view on a hyperbolic tiling from the game hyperrogue. The second image shows the hyperbolic rooms of hyperbolica. The third image shows a hyperbolic grid from hyperNom. The fourth image shows the hyperbolic world of HypVR, the fifth shows a greyish variant. The sixth image shows the hyperbolic world of HypVR2.
    }
    \label{fig:otherGames}
\end{figure*}

Virtual environments including hyperbolic spaces have been included in a few previous works, see Figure~\ref{fig:otherGames}.
Those are set in two-dimensional hyperbolic space~$\mathbb{H}^2$, three-dimensional hyperbolic space~$\mathbb{H}^3$, or a combination~$\mathbb{H}^2\times\mathbb{E}$, where the floor is a two-dimensional hyperbolic space and the height-dimension is regular Euclidean.
Skrodzki~\cite{skrodzki2021illustrations} provides a general review of illustrations of non-Euclidean geometry in VR.

\emph{HyperRogue} is a game set on the hyperbolic plane~\cite{koczynski2017hyperrogue}.
The player is shown a top-down view of the game world, rendered using the Poincaré disk model, see Figure~\ref{fig:Hyperrogue}. 
It is a rogue-like, turn-based game in which the player moves the character from one tile to a neighboring tile every turn or executes an attack on an adjacent tile.
\emph{HyperRogue} gives the player an idea of what it is like to navigate tilings of the hyperbolic plane.
While a VR mode for \emph{HyperRogue} exists using~${\mathbb{H}^2\times \mathbb{E}}$, it only allows for teleportation as a locomotion technique.
This keeps the user static and does not use the move area to connect physical and virtual space.

\emph{Hyperbolica} is a game set in a continuous three-dimensional hyperbolic space~\cite{hyperbolica}. 
It is a 3D game that supports VR, see Figure~\ref{fig:Hyperbolica}. 
While \emph{Hyperbolica} is set in~$\mathbb{H}^3$, it uses~${\mathbb{H}^2\times \mathbb{E}}$ for the underlying physics.
In the game, movement by walking is limited and locomotion is realized via teleportation for larger distances.
This disconnects the user and hinders full immersion.
\emph{Hyperbolica} illustrates effects such as holonomy when the user walks around but does not use it to connect physical walking with movement in the virtual world.

The task in the VR experience \emph{Hypernom}---see Figure~\ref{fig:HyperNom}---is to `eat' all cells of a regular 4D polytope, which is radially projected into the three-sphere~$\mathbb{S}^3$~\cite{hart2015hypernom}.
The user moves through the environment by rotating their head in the physical space.
Based on the unit quaternions representing the headset's three-dimensional rotation, the player's physical movements are translated to moves in the virtual space.
While this provides a high degree of immersion, no movement with the entire body is possible and the experience is limited to `eating` one polytope.

\emph{HypVR}~\cite{hart2017non-euclideanI,hart2017non-euclideanII} is a hyperbolic space simulator incorporating several aspects that we will use within \emph{Holonomy}.
In \emph{HypVR}, users can explore both~$\mathbb{H}^3$ and~$\mathbb{H}^2\times \mathbb{E}$, see Figures~\ref{fig:HypVR} to~\ref{fig:HypVR2}.
However, it shows a continuous space the user cannot fully explore.
Movement in the world is possible and the user can utilize holonomy similar to the usage in our environment---see below---to access the entire hyperbolic disk or sphere.
However, the spatial limitations of the physical space are not reflected in the virtual display, causing a disconnect between virtual movement possibilities and physical walking limitations.
As all user movement happens in a pre-rendered cube, the population of the worlds by different objects becomes impossible, as shown by Skrodzki~\cite{skrodzki2021illustrations}.

As stated in Section~\ref{sec:Introduction}, our virtual environment overcomes the limitations of previous work.
In particular, we enable users to freely walk for infinite exploration, eliminating the need for teleportation.
Furthermore, we provide a populated space that can be used for various application scenarios.
\section{\emph{Holonomy}---the virtual environment}
\label{sec:HolonomyTheVirtualEnvironment}

In this section, we describe several design choices made in the development process of  \emph{Holonomy}.
First, we assume the move area to be roughly a square of at least~${3\times 3}$ meter unobstructed physical space, subdivided into nine~${1\times 1}$ meter cells, confer the floor in Figure~\ref{fig:teaser}, right.
The virtual environment is designed with these proportions in mind and it currently does not include support for other grid sizes.
The user's walking in this physical space is mapped to movement in the virtual hyperbolic space.

We have chosen to use a hyperbolic square tiling with five squares meeting at every vertex.
This has two major benefits over other tiling options. 
First, squares map nicely to Euclidean space.
Hence, the~${3\times 3}$ squares in the move area can be easily aligned with the virtual ones. 
Second, five meeting squares is the smallest number to exhibit hyperbolic geometry.
Thus, the user can more easily perceive the holonomy effect. 
With more squares meeting at a vertex, the user has to run through several extra rooms around a vertex to return to the original square.
This would make it less likely that they actively experience holonomy, as the phenomenon cannot be perceived immediately but only by keeping track of movements.

\begin{figure}
    \centering
    \includegraphics[width=1.\linewidth]{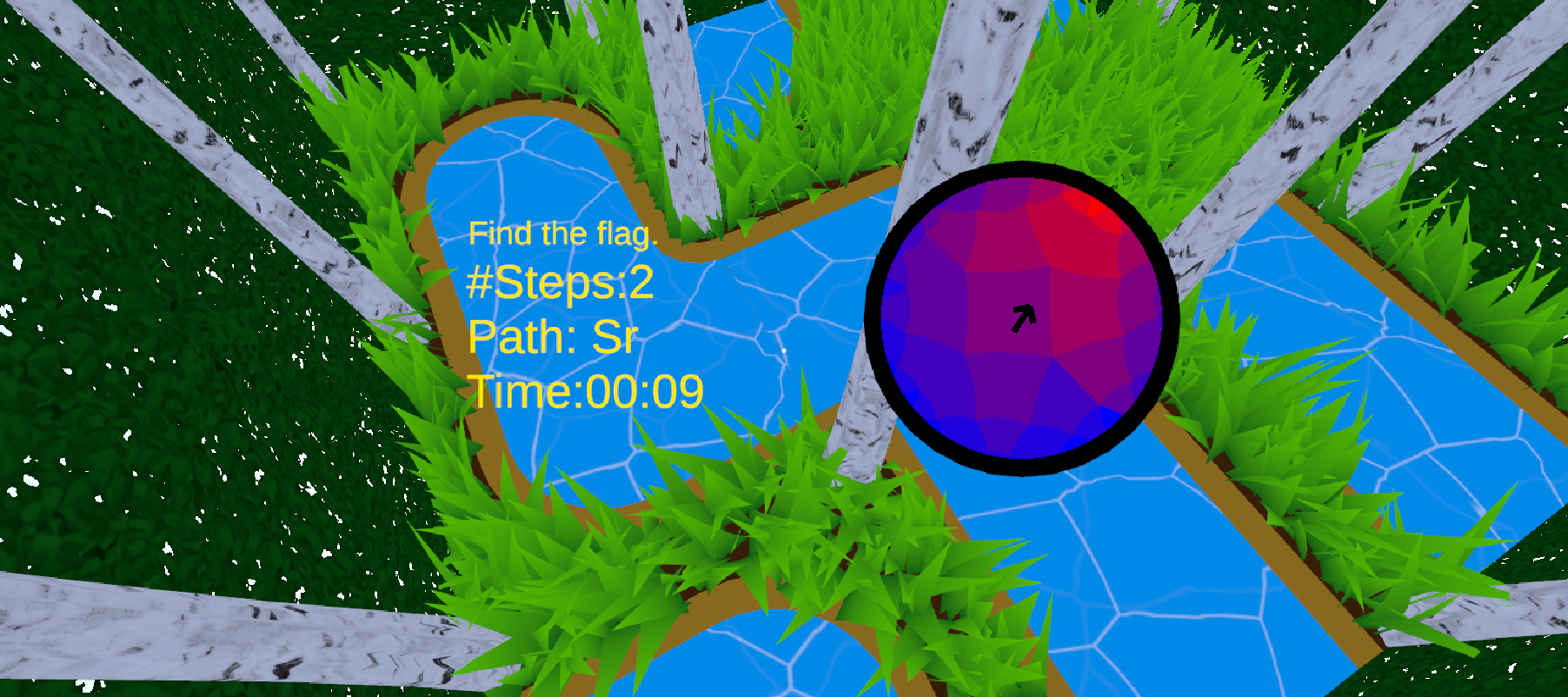}
    \caption{
        The user interface shows the current status and the mini-map, helping with navigation.
    }
    \Description{
        An image of the game world with the user interface. A ${3\times 3}$ grid can be seen separated with trees at the corner. The floor consists of grass and small rivers. 
        Hedges are visible at the edges of this ${3\times 3}$ grid.
        On the left of the image yellow text is shown, in 4 lines stating: Find the flag, \#steps:2, Path: Sr, Time:00:09. 
        On the right, a circular mini-map is visible. 
        At the center of the mini-map, a black arrow can be seen pointing to the top right. 
        On the mini-map, several areas are visible. 
        Each is uniformly colored with a color on a total gradient from the top right area being red and the bottom left area being blue.
    }
    \label{fig:HolonomyUI}
\end{figure}

As only~${3\times 3}$ squares in the move area are physically accessible, we also only render these accessible squares virtually.
While the mini-map indicates that the user is moving in an infinite environment, see Figures~\ref{fig:HolonomyUI} and~\ref{fig:minimaps}, limitations are shown in the virtual environment.
As an application scenario, we chose to render a virtual, hyperbolic forest, therefore, the limitations are rendered in the form of hedges, see Figure~\ref{fig:vrworld}. 
These coincide with the move area's boundaries, providing a safe exploration environment.

To test the capabilities of our virtual environment, we confront users with objectives to follow.
One objective prompts the users to navigate to certain squares.
Usually, these squares are initially out of reach, as they are outside the accessible~${3\times 3}$ grid.
However, applying the holonomy concept shifts different squares into the accessible area.
Users can visit any cell in the hyperbolic grid by continually applying these transformations, including their objective.

\begin{figure}
    \centering
    \includegraphics[width=1.0\linewidth]{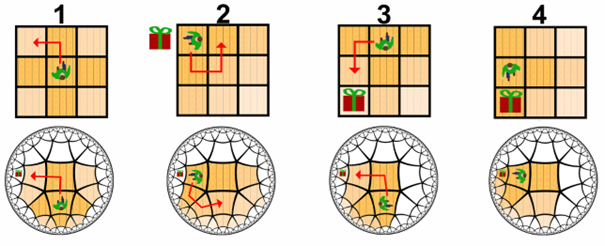}
    \caption{
        The top row indicates the user walking in the move area. 
        The bottom row indicates how the user and the objective move within the virtual hyperbolic space, those squares accessible by the user are shaded.
    }
    \Description{
        2 rows and 4 columns are shown each column being labeled from 1 to 4 above the column. the top row has 4, ${3\times 3}$ squared brown colored grids. At the center of the first grid, a character is shown with a red arrow indicating where the character is located next. At grid number 2 the character is in the top left. A gift box is shown outside the grid to the left of the character. In grid 3 the character is in the top middle square and the gift is in the bottom left. In grid 4 the character is in the middle left right above the gift. In the bottom row, 4 hyperbolic disks are shown. On these disks, 9 areas are colored brown. On the first disk, the character is in the area below the center and a red arrow shows to what area the character will move next. To the left of the topmost brown area, a gift box is shown in a white area. In the second disk, the character is in the area to the right of the gift box which is still in a white area. In the third disk, the character is again below the center and the gift box is now in a brown area. In the fourth disk, the character is to the right of the gift box which is located in a brown area.
    }
    \label{fig:hyperbolic_walker_2}
\end{figure}

Consider Figure~\ref{fig:hyperbolic_walker_2} for an example of applying holonomy to reach an objective.
Here, the objective for the user is to access the square marked with the icon showing a present.
Initially, this square is out of reach.
By walking a full circle in the move area, the holonomy concept causes a discrepancy between the physical and virtual position of the user as illustrated in Figure~\ref{fig:HyperbolicWalker}.
This shifts previously inaccessible virtual squares into the accessible area for the user to reach them.
Repeated application of this principle lets the user reach any position in the infinite virtual hyperbolic plane, while only walking within their~${3\times 3}$ move area.

To support the users in fulfilling an objective, we provide them with navigational tools.
Namely, we use a Poincaré disk model to show the user a larger part of the virtual, hyperbolic world expanding from the currently accessible area.
Each square in the mini-map is colored according to its distance to the closest objective, where red colors are close to (hot) and blue colors are far from (cold) the goal, see Figure~\ref{fig:minimaps}, second from right.
This coloring represents a form of an unobstructed Manhattan distance on the tiling.
Additionally, we display the shortest path the user needs to walk to return to a reasonable distance to the objective whenever they stray away too far.
This prevents users from becoming completely lost, see Figure \ref{fig:minimaps}, rightmost.

For the positioning of the mini-map, we offer two different views: One in the HUD and one on the VR controllers, as a map to be held by the users.
In an initial user study comparing these different views, no link between them and the navigational performance was found~\cite{jochems2023mini-map}.
However, different users did prefer one mode over the other.
Hence, we offer both map views as a choice.

\section{Challenges and solutions}
\label{sec:ChallengesAndSolutions}

Implementing the environment with the features described above poses several technical challenges.
In the following, we list these and our respective derived solutions.
We developed the environment within the \emph{Unity} game engine and tested it in VR on an HTC \emph{Vive} and a Meta \emph{Quest} headset.
Additionally, the environment offers a pancake mode to explore with a screen and a controller.
However, the main feature of having the user use physical walking as a control mode to navigate the virtual environment naturally only comes within the VR setup.

Note that all challenges contribute to the usability of this versatile environment.
As it is based on the unfamiliar geometry of hyperbolic space, indicating direction helps new users to navigate the virtual environment.
A correct rendering allows us to capture and visualize the effects of hyperbolic geometry.
A solution for the population of the virtual environment makes it a versatile tool, which can then be used to, for instance, represent information, provide a space for digital co-creation, or play games.
We will discuss these aspects in detail in the following.

\subsection{Indicating directions}

Note that for a given objective, the target tile might not only lie outside the currently accessible area but even outside of the tiles currently displayed on the mini-map.
Thus, it is necessary to indicate the general direction of such objectives to the user.
To do this, the user's and the target tile's locations are converted into two points in Minkowski hyperboloid coordinates~\cite[Chapter~3.6]{kinsey2011geometry}. 
Using these coordinates, the direction can be found by taking the~$x$ and~$y$ coordinates of both points and computing the angle between these two points in two dimensions. 
Since the angle is only used as a visual indicator on the mini-map, it does not have to be very precise. 
Therefore, the usual mathematical inaccuracy of this process is not an issue.

\subsection{Rendering}
\label{sec:Rendering}

\begin{figure}
    \centering
    \includegraphics[width=0.8\linewidth]{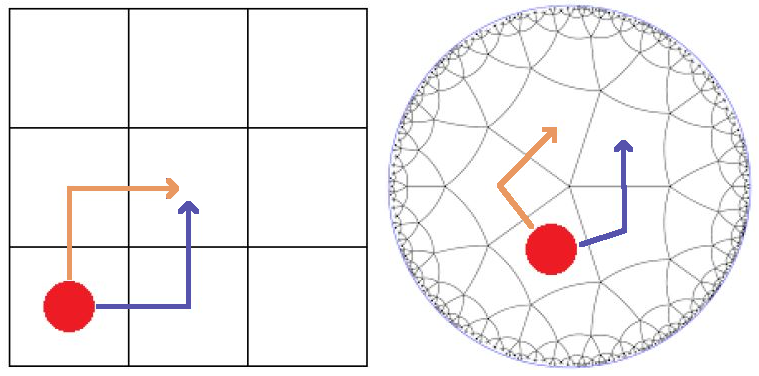}
    \caption{
        Moving along the blue and orange arrow in the physical space leads to the same position in the physical space (left), but not to the same position in the virtual space (right).
        Replicated with permission~\cite{slotboom2023rendering}.
    }
    \Description{
        On the left a ${3\times 3}$ grid is visible. In the bottom left square a red dot is drawn. An orange arrow goes from the red dot up and to the right. A blue arrow goes from the red dot right and up. Both arrows end in the center square. On the right, a hyperbolic disk is shown. A red dot is shown in one of the areas. An orange arrow goes from the red dot to the top left and then the top right. A blue arrow goes from the red dot to the top right and then the top left. Both arrows end in different areas.
    }
    \label{fig:HyperbolicViewDifference}
\end{figure}

The process of rendering a hyperbolic tiling using an Euclidean render engine presents challenges arising from the need for diverse tiles to coexist within the same Euclidean space, as depicted in Figure~\ref{fig:HyperbolicViewDifference}.
Namely, two different tiles in the hyperbolic world can occupy the same coordinates in the Euclidean representation.
Following the orange or blue arrow, respectively, lets the users stand in the same physical square, but has them end up in different virtual tiles.
Going around the vertex via the left (orange arrow) should thus show a different tile than going around via the right (blue arrow).
This can be seen in Figure~\ref{fig:vrworld}, where the red flag is visible around the right view of the tree, but not around the left.

\begin{figure}
    \centering
    \includegraphics[width=0.8\linewidth]{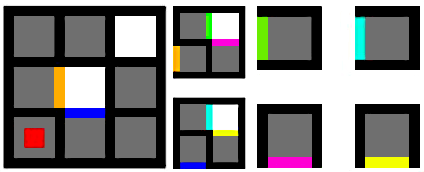}
    \caption{
        Splitting the~${3\times 3}$ grid into subgrids, with the portals to connect them shown as colored edges.
        Replicated with permission~\cite{slotboom2023rendering}.
    }
    \Description{
        On the left, a ${3\times 3}$ grid is shown. In the bottom left of this grid, a red square is shown. The middle and the top right square are colored white and the rest is colored dark grey. The line splitting the middle square and the square to the left of it is colored orange and the line to the bottom of it is colored blue. In the middle of the image 2, ${2\times 2}$ grids are visible. In the top grid, the bottom left square has an orange line on the left. The top right square is white and to the left, the line is lime green and the bottom is pink. In the bottom grid, the bottom left square has a blue line beneath it. The top right square is colored white and has a cyan line to the left and a yellow line to the bottom. On the right of the image, 4 separated squares are shown. The top left square has a lime green line to the left. The top right square has a cyan line to the left. The bottom left square has a pink line to the bottom. The bottom right has a yellow line to the bottom.
    }
    \label{fig:portals}
\end{figure}

Our first solution to this problem was to divide the grid into \emph{sub-grids} and to establish rendering \emph{portals} that connect the sub-grids to the primary grid, see Figure~\ref{fig:portals}.
The view according to the orange arrow of Figure~\ref{fig:HyperbolicViewDifference} would thus show a portal rendering a~${2\times2}$ geometry somewhere far away in the scene, shown in the second column of Figure~\ref{fig:portals}.
The same would happen for the blue route.
As the phenomenon is also present in the~${2\times 2}$ grids, these would be split again, leading to four more portals and geometries, see the right of Figure~\ref{fig:portals}.
Overall, depicting the hyperbolic space in this fashion calls for rendering not one, but seven views.
This quickly becomes a performance issue in VR, where high frame rates are required for a comfortable user experience.

To overcome this problem, instead of using expensive render textures, we use \emph{stencil polygons}~\cite{Neerdal2020} to create the portals. 
Stencil polygons are transparent objects that write a value to the stencil buffer, which can be read to determine whether an object placed behind the stencil should be rendered.
This allows an object to be placed in the same position as another while still only being visible when looking through one stencil polygon and not the other---that is, being visible through the orange but not the blue stencil in the context of Figure~\ref{fig:portals}.
This reduces the render load drastically as only one render queue is necessary, instead of one render queue per portal.
For all technical details of this implementation and a benchmark of time game in rendering, we refer to a corresponding tech report~\cite{slotboom2023rendering}.

\subsection{Populating the virtual space with objects} 
\label{sec:PopulatingTheVirtualSpaceWithObjects}

An empty virtual environment is not of much use, as all tiles look the same, aside from their colors, see Figure~\ref{fig:otherGames}, four rightmost images.
This also hinders effective navigation due to the absence of meaningful landmarks.
Only population with objects representing, for instance, data, introduces the possibility of integrating the environment in a specific application.
It will help to differentiate the different tiles.

The general challenge for populating is, however, regarding previous work, that adding ``such objects would call for tremendously different rendering algorithms''~\cite[p.~212]{skrodzki2021illustrations}.
Specifically, recall the exponential growth of a hyperbolic disk regarding its radius, which can lead the user to potentially visit thousands of distinct tiles within ten steps from the origin.
Manual design becomes infeasible for such numbers of tiles.
Therefore, we employ a \emph{Wave Function Collapse} (WFC) algorithm~\cite{mxgmn} to procedurally populate the virtual environment.

\begin{algorithm}
	\caption{Wave Function Collapse (WFC)}
	\begin{algorithmic}[1]
		      \While{an uncollapsed tile exists}
			     \State Tile $T \gets$ GET any tile with lowest entropy
			     \State Object $O \gets$ Collapse $T$ to show any of its possible objects
			     \For{all neighbors of $T$}
				    \State Remove objects incompatible with $O$
			     \EndFor
		      \EndWhile
	\end{algorithmic}
	\label{alg:wfc}
\end{algorithm}

The specific algorithm we implement is similar to the WFC algorithm examined by Karth et al.~\cite{karth2017WFC} and presented as Algorithm~\ref{alg:wfc}.
We choose it, as it has been proven to have great potential for generating populated environments~\cite{cheng2020automaticgeneration}. 
Our algorithm first collects all tiles around the user. 
It is then executed on all uncollapsed tiles, that is, those tiles without an object.
While we do not use a grid, we can get the neighbors per tile based on the underlying graph structure, see Figure~\ref{fig:spanning}. 
This enables us to impose constraints on the neighbors of each tile.

Once a tile has collapsed into an object, it will impose constraints specific to that object onto its neighbors. 
Constraints can limit the set of possible objects within a tile or force neighboring objects to connect in a certain orientation, like a jigsaw puzzle. 
These objects together form a more immersive environment for the user than an empty world~\cite{ImmersionRijsdijk2023}. 

Constraint solving conflicts could prevent stock WFC to finalize, and good implementations aim to mitigate them as much as possible.
They arise whenever a tile has no possible objects to collapse into, as too many constraints have been imposed upon it by its neighbors. 
We reduce the chances of a conflict by collapsing tiles according to increasing entropy~\cite{karth2017WFC}. 
If one occurs, we backtrack.

Using this algorithm creates an initial population of the virtual environment.
See Section~\ref{sec:ProofOfConcept} for an illustration within an application example.
Naturally, this initial population can be manually edited by a designer, who can easily add, move, or change specific highlights or points of interest, without going through the laborious task of designing ``all tiles'' for the environment.

\subsection{Navigational Guidance}
\label{sec:Navigation}

Since navigation in a hyperbolic space can be confusing, it is convenient to provide ways to guide the user in the desired direction.
This also aids the users in learning the mechanics of the hyperbolic virtual environment. 
One way is to show them the shortest path to the objective. 
Yet, the computation of the shortest path has to take into account both the currently accessible tiles and the entire hyperbolic virtual space.
Deriving the shortest path under only one of these circumstances is relatively simple. 
However, combining both of them results in a more complex problem that requires a new way of finding the shortest path.
The main obstacle here is that the user's orientation has to be taken into account as it shifts the accessible tiles, confer Figure~\ref{fig:hyperbolic_walker_2}.

\emph{Holonomy} models the problem as a graph so existing graph traversal algorithms can be used. 
The attributes used to define each node encompass all possible locations in the virtual space, the move area, and the user's respective rotation. 
They allow a node to calculate what its neighboring nodes are lazily, without requiring any past knowledge. 
Storing a graph of all locations with a decent radius around the starting point is not feasible because of the exponential growth of the hyperbolic space regarding the distance from the origin. 
Therefore, lazy generation of the graph is essential. 

We use the A* algorithm as the shortest path algorithm on our graph. 
The A* algorithm is a variation of the Dijkstra shortest path algorithm that adds a heuristic to the traversal process to guide the node expansion. 
The performance of A* depends on the heuristic that is used~\cite{Hart1968}. 
The heuristic we employ is the length of the shortest path in the two-dimensional hyperbolic space, without regarding the currently accessible tiles.
This is a relaxed version of our problem and is thus admissible. 
It guarantees that A* returns an optimal path~\cite{Hart1968}. 

To ensure quick user feedback, \emph{Holonomy} keeps the runtime low by switching the shortest path algorithm depending on the user's distance to the nearest objective.
When the user is close to the objective it uses the A* algorithm. 
If the user is far from the nearest objective, it switches to an \emph{anytime} variant of the A* algorithm. 
This variant tries to approximate the start of the shortest path but does not guarantee that the returned path will end at the objective. 
However, complete paths are unimportant for larger distances, as the user cannot see further than three tiles away from their position on the mini-map, confer Figure~\ref{fig:minimaps}, rightmost. 
This variant still uses the same A* algorithm at its base but has an extra parameter representing a time limit. 
The algorithm keeps track of the path that gets closest to the objective according to the heuristic. 
When the time limit is reached and the algorithm has not found the optimal path, it returns the saved path. 
In our preliminary experiments, for a returned path of length~$n$, the algorithm has always returned the first $n$ steps of the optimal path~\cite{Snellenberg2023}.

A user of the environment might want to navigate to several objectives within the environment.
Calculating the shortest path between multiple locations in \emph{Holonomy} is not as straightforward as a normal implementation of the \emph{Traveling Salesperson Problem} (TSP). 
The shortest path algorithm does not consider what rotation (both in the virtual environment and the physical move area) and position in the move area the user reaches the location in. 
However, those properties influence the paths that can be taken from that location.
Consequently, if the shortest path from point~$A$ to point~$B$ goes through~$C$, then the shortest path to~$C$ is not necessarily contained in the shortest path from~$A$ to~$B$. 
Thus, the described shortest path algorithm violates the optimal substructure property which is a requirement to use existing TSP solvers.
This issue can be solved by modeling the problem as a set TSP problem, where all possible states of a certain objective location constitute a set. 
In the case of \emph{Holonomy}, each location is linked to~$144$ states since the user can be in one of~$9$ move area states and one of~$4$ rotations in both the move area and the hyperbolic plane. 
However, the size of a set can be reduced to~$9$ by taking advantage of the symmetries present in the environment. 

\section{Proof of Concept: A navigation game}
\label{sec:ProofOfConcept}

\begin{figure}
\centering
    \includegraphics[width=1.\linewidth]{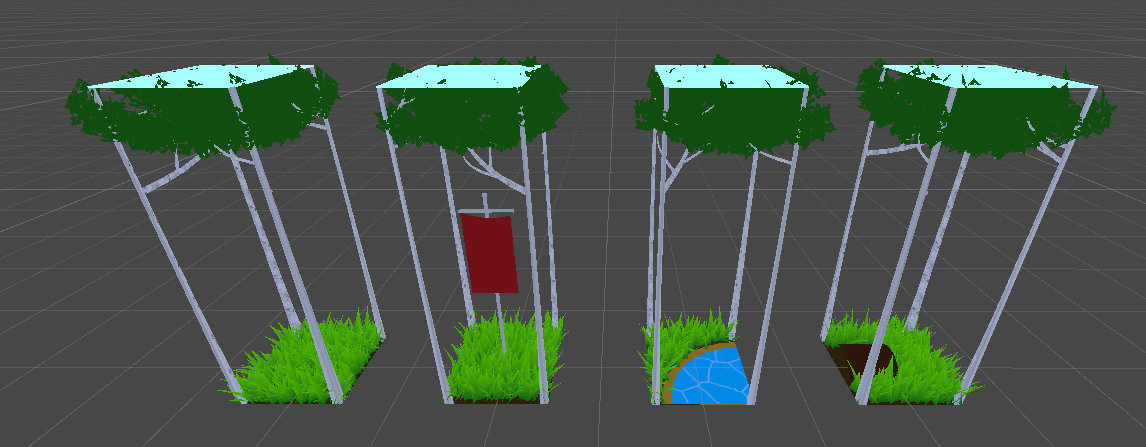}
    \caption{
        Four examples of natural objects used to populate the world. 
        The tree trunks in each corner of each square serve as natural separating pillars. 
    }
    \Description{4 tiles from the game world are shown. All have trees on the corners of the squares. The first tile has grass on the floor. The second tile has grass with a flag. The third tile has grass with a curved river. The fourth tile has half a log and grass.}
    \label{fig:tiles}
\end{figure}

To illustrate the capabilities of our virtual environment, we implemented a VR treasure hunt game.
The game is set in a forest and asks users to navigate to certain objectives.
To reach these, the users get varying degrees of support.
We decided to use simplified natural-looking objects to populate the world, examples of which are shown in Figure~\ref{fig:tiles}.

\begin{figure}
\centering
    \includegraphics[width=1.\linewidth]{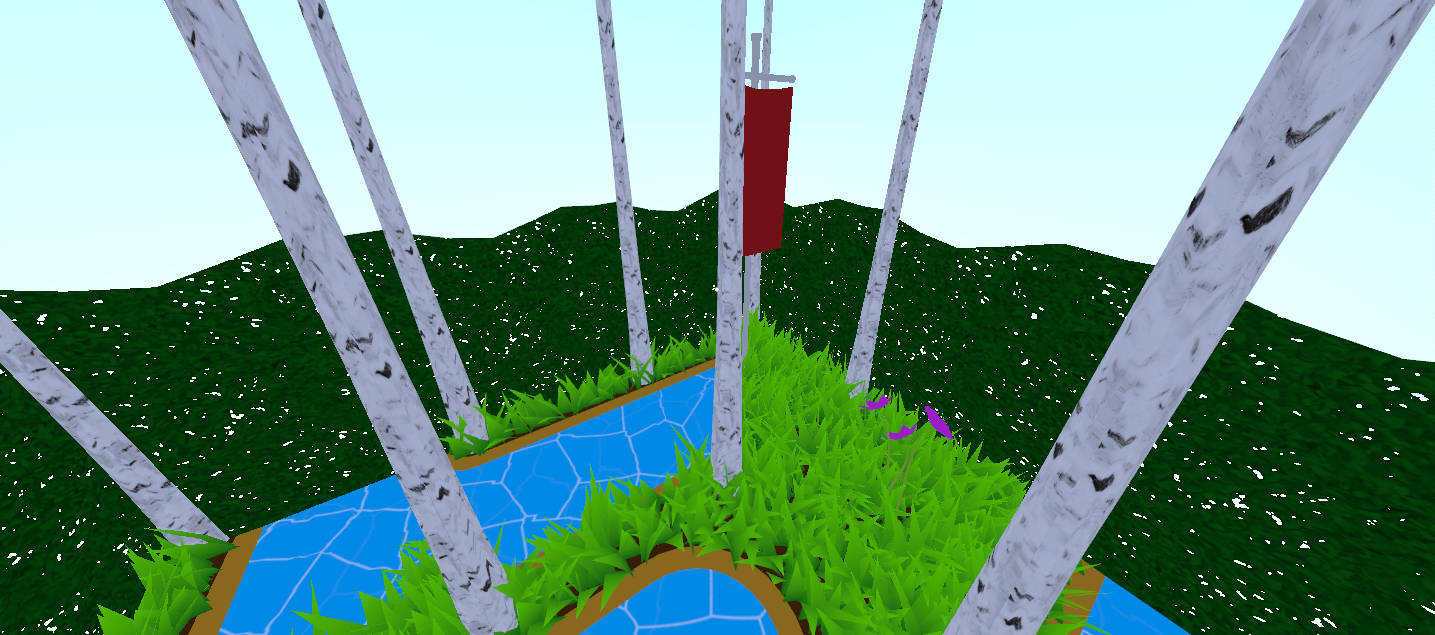}
    \caption{
        An Example of the world as seen in VR.
    }
    \Description{
        An image of the game world is shown with the player in the bottom right looking towards the top right square. A ${3\times 3}$ grid is visible on the corners trees are visible. On the edge of the grid, a hedge is visible. On the floor grass, rivers, and flowers are visible. A flag is visible in the top right square.
    }
    \label{fig:vrworld}
\end{figure}

Within this forest environment, the objects serve as orientation landmarks.
Aside from flowers occupying a single tile, as shown in Figure~\ref{fig:teaser}, logs occupy two neighboring tiles, while creeks continue over several tiles.
This contributes to highlighting the hyperbolic effects of the environment, as creeks and logs suddenly vanish behind one side of a tree, confer Figures~\ref{fig:teaser} and~\ref{fig:vrworld}.
Furthermore, the trees we place at the vertices where five squares come together serve as natural separating pillars with diverging views around their left or right side, see Figure~\ref{fig:vrworld}.

\begin{figure}
\centering
    \includegraphics[width=8.5cm,height=9cm,keepaspectratio]{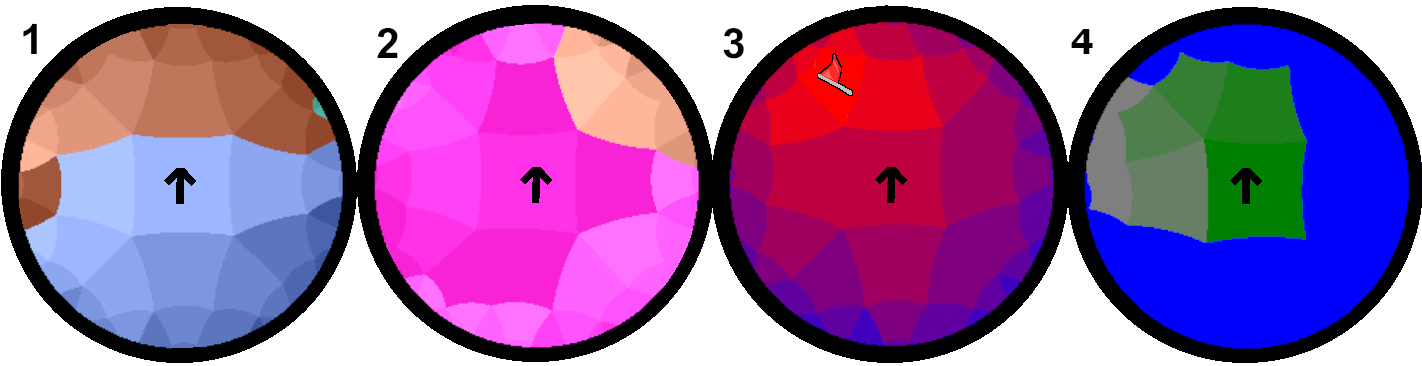}
    \caption{
        Images 1 and 2 display two different cases of biome propagation in the hyperbolic grid. 
        Image 3 shows an instance of hot-cold coloring in the minimap. 
        Image 4 shows an instance of the shortest path to the closest objective.
    }
    \Description{
        Four hyperbolic disks are shown. The first shows the left quarter to be pink while the top, right and bottom are colored salmon orange. The second shows the top right being pink, the top left light blue and the rest is colored turquoise. The third has a single area containing lines on the top left from this area a gradient is drawn around it where the closest areas to the striped area are colored red and areas the furthest away are colored blue. The areas in between follow a gradient from red to blue depending on how close they are to the striped area. The fourth disk is blue with a green path being shown in the bottom left.
    }
    \label{fig:minimaps}
\end{figure}

In the WFC algorithm, we can introduce a hierarchy~\cite{alaka2023hierarchical} to group similar objects. 
Here, we limit the hierarchy to one high-level layer on top of the objects, representing \emph{biomes}.
We use a simple propagation algorithm to generate biomes, the result of which is shown in the left two images of Figure~\ref{fig:minimaps}: If a square has no biome, we assign one and set its propagation depth to a value greater than zero.
If a square collapses and its depth is greater than zero, it propagates its biome to neighbors without a biome and decreases its propagation depth by one.
The type of biome then determines the possible objects for the tile.

The navigation aspect discussed in Section~\ref{sec:Navigation} provides a tutorial tool for users to learn how to navigate in the virtual environment.
However, it also allows the users to compare the number of steps taken toward the goal with the optimal number represented by the shortest path.
It motivates them to improve their navigational skill and to re-play a level.

\begin{figure}
    \centering
    \includegraphics[width=0.8\linewidth]{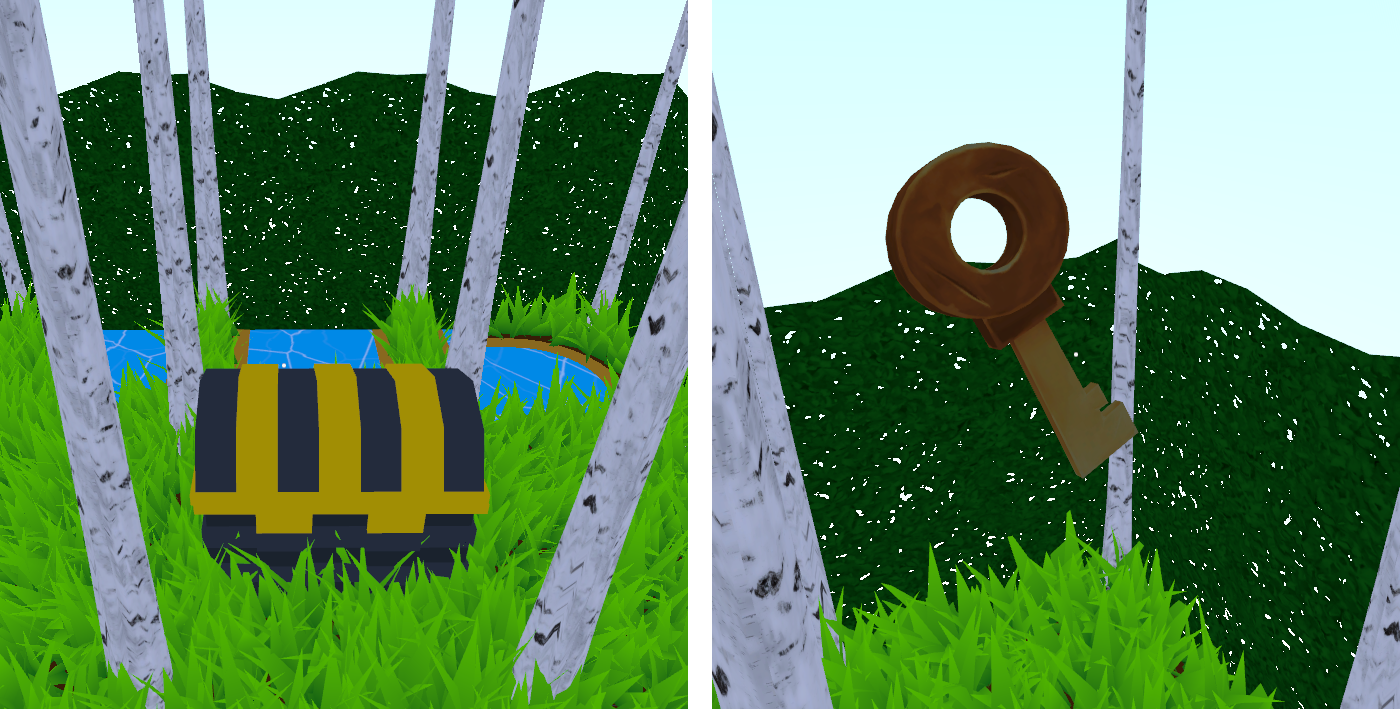}
    \caption{Left: a chest; Right: the required key.}
    \Description{On the left image a chest is visible sitting in the grass on the center square of the game world. On the right a floating key is visible in a corner of the game world the floor is grass and the walls are hedges.}
    \label{fig:ChestKey}
\end{figure}

Two game modes are currently implemented.
The first asks a user to reach a flag, as shown in Figures~\ref{fig:teaser},~\ref{fig:tiles}, and~\ref{fig:vrworld}.
This serves as a first-stage navigational task, aided by navigational cues.
The second mode requires the user to collect several keys and bring them to a chest, see Figure~\ref{fig:ChestKey}.
This requires finding the optimal way including several objectives, along a TSP problem as discussed in Section~\ref{sec:Navigation}.
To learn the basic functionality of the game, users are walked through a tutorial level.\footnote{Find a video of the tutorial with explanations here: \url{https://youtu.be/wL37nyJ_ECM}.}

We let several users interact with our virtual environment based on this proof of concept.
One user study with~23 participants explored the impact of the procedurally generated environment on the user's immersion~\cite{ImmersionRijsdijk2023}.
This study reports an increasing immersion due to the population of the environment.
A second study with~30 participants investigated the placing of the mini-map on either the HUD or the controller of the user~\cite{jochems2023mini-map}.
The findings show that the optimal position mostly depends on the preference of the individual user.
Furthermore, the prototype was shown at the \emph{Dutch Game Gardens} event, where feedback from~12 more users was gathered qualitatively based on the following questions:
\begin{enumerate}
    \item What is your spatial perception of the environment? How does it feel?
    \item How did you approach the navigation task?
    \item When navigating: how did the mini-map help (or not) in completing this task?
    \item When navigating: how did the colored poles help (or not) in completing this task?
    \item What applications do you see for this game?
    \item What other suggestions or comments do you have?
\end{enumerate}
All user groups were comprised of people with (extensive) prior VR experience and people that never used VR before.

From these interactions, aside from the validated findings reported~\cite{ImmersionRijsdijk2023, jochems2023mini-map}, we made several anecdotal findings, that warrant to be followed up on.
First, it does take the users some engagement with the platform to overcome their initial confusion, which is rooted in the underlying hyperbolic geometry.
Three participants from the qualitative study described the environment as ``confined'' or ``claustrophobic''.
One of the participants stated regarding the first question: ``It feels claustrophobic and a bit disorienting or even frightening, but also really intriguing and gives a great curiosity for exploration.''
Three other participants described their spatial perception of the environment as ``good'' with one user specifying: ``Intriguing as soon as I saw the effect of items 'around the corner'.''

In contrast to the initial spatial perception, controlling movement in the environment came naturally to all participants, as they walked and moved to explore it.
Most noteworthy is that due to the natural walking movements, no participant had to abort testing the prototype due to VR sickness.

Navigation in the environment seems to work well for some users, others need more time.
However, most players improve their navigational abilities through a few initial interactions.
Overall, the proof of concept showed the feasibility of navigating and exploring a potentially infinite virtual world via physical movement in a restricted move area.

\section{Conclusions: Impact and outlook}
\label{sec:ConclusionsImpactAndOutlook}

We have presented  \emph{Holonomy}, a virtual environment based on the mathematical concept of hyperbolic space.
Unlike other virtual environments, it allows users to explore a potentially infinite virtual space solely by walking in a confined physical move area, without the need for artificial locomotion such as teleportation.
We described how to overcome several technical challenges related to rendering and populating the space and how to compute the shortest path to an objective to initially guide the user before they master navigation in the space.
In addition, we give two examples where the application of such an environment has an impact.

First, hyperbolic geometry, leaving behind the familiar Euclidean principles, presents a significant challenge for learners due to its unintuitive nature~\cite[Chapter~V]{armand2019relevance}. 
Aspects such as holonomy, the concept of how a geometric object evolves as it moves along a curve, can be difficult to grasp without tangible visualizations or hands-on experiences~\cite[p.~79]{shapovalova2020hyperbolic}. 
Traditional educational methods often struggle to convey these abstract ideas adequately. 
The virtual environment presented in this paper offers a promising base for building future learning techniques.
We hope learners can develop a deeper intuition for the underlying principles by navigating through hyperbolic spaces and observing how shapes and objects behave within these environments.
The educational value of our platform was already exemplified by incorporating a corresponding task in the 2023 edition of the ``Mathekalender,'' a project for mathematics education\footnote{
    Find the task formulation here: \url{https://www.mathekalender.de/wp/calendar/challenges/2023-10-en/}.
}.
Thus, we see great potential impact in implementing our platform within an educational context.

Second, integrating representations of hyperbolic geometry into VR environments offers a novel approach to studying navigational challenges in psychological research. 
Unlike traditional Euclidean maps, which study how participants are accustomed to navigating, hyperbolic spaces present entirely new and unfamiliar landscapes.
By immersing participants in hyperbolic environments through VR, researchers can create unique navigational challenges that require individuals to adapt their spatial reasoning and cognitive strategies. 
Observing participants in their learning process on navigating might offer insights into how navigational strategies are formed and what role human cognition and perception play in these strategies.
In these experiments, the immersive nature of VR allows for controlled setups with various hyperbolic geometries, enabling researchers to investigate how different features, such as tiling variations, influence navigational performance and cognitive load. 
This opens avenues for studying spatial cognition and advancing our understanding of human navigation in unfamiliar and complex environments.

Finally, considering these and other potential use cases, our virtual environment has technical challenges that remain unanswered. 
One example is the rendering of light. 
While hyperbolic space is a well-described mathematical concept, light propagation in a virtual environment context has not been explored. 
Regarding our setup, spotlights in one square tile can affect neighboring tiles in non-euclidean ways, requiring a novel technical implementation currently unsupported by the rendering pipelines.

Another challenge is the incorporation of further physical boundaries.
In our proof of concept, the physical boundaries of the move area are rendered as hedges within the virtual space.
However, there are no measures in place that prevent the user from walking through these virtual boundaries, leaving the move area.
Placing physical hedges, or dummies that mimic such, on the boundary of the move area would allow the users to reach out to hedges they see in the virtual world and touch them physically.
The same could be done for the trees on the vertices between square tiles.
This would contribute to a significantly stronger connection between the virtual and physical worlds, as suggested in previous research~\cite{melo2020multisensory}.
All these challenges are left as future work.

\bibliographystyle{ACM-Reference-Format}
\bibliography{bibliography}


\end{document}